# High-Temperature-Resilient Hyperbolicity in a Mixed-Dimensional Superlattice

Jason Lynch[1, †], Tzu-Yu Peng[2, 3, †], Jing-Wei Yang[2, 3], Ben R. Conran[4], Bongjun Choi[1], Cindy Yueli Chen[5], Zahra Fakhraai[5], Clifford McAleese[4], Yu-Jung Lu[2, 3, *], Deep Jariwala[1, *]

[1] Department of Electrical and Systems Engineering, University of Pennsylvania, Philadelphia, Pennsylvania 19104, United States

[2] Research Center for Applied Sciences, Academia Sinica, Taipei 11529, Taiwan

[3] Graduate Institute of Applied Physics, National Taiwan University, Taipei 10617, Taiwan

[4] AIXTRON Ltd, Swavesey, Cambridge CB24 4FQ, UK

[5] Department of Chemistry, University of Pennsylvania, Philadelphia, PA 19104, USA

[*] Corresponding Authors: dmj@seas.upenn.edu , yujunglu@gate.sinica.edu.tw

[†] These authors contributed equally to this work

## Abstract

Hyperbolic superlattices are used for sub-wavelength focusing, cloaking, and optical thermal management. Typically, these superlattices are constructed of layers of noble metals and insulators. Despite these systems displaying excellent optical performance, the poor thermal stability of noble metals prevents their application in high-temperature environments. Instead, CMOS-compatible transition-metal nitrides are often substituted for noble metals in plasmonic systems since they have high thermal stability at the expense of optical properties. Here, we fabricate hyperbolic titanium nitride (TiN)/hexagonal boron nitride (hBN) superlattices with 3D-2D interfaces. The mixed-dimensional nature of the interfaces prevents atoms from diffusing across the interface at high temperatures. The hyperbolicity of the superlattice is found to be unaffected by annealing at high temperature (800 °C for 10 hrs), and TiN/hBN is found to have a larger hyperbolic figure of merit than similar superlattices.

## Introduction

Metamaterials facilitate optical phenomena that are rarely observed in natural materials such as sub-wavelength focusing[1,2], cloaking[3,4], and perfect absorbers[5–7]. These properties emerge from the subwavelength patterns of the metamaterial, and they have enabled high-performance, ultra-small optical devices. Although metamaterials can also be based on Mie[8] and exciton[9] resonances, many metamaterials rely on collective oscillations of free charges (plasmons[10]) in metals to confine light on the nanometer-scale. Plasmons are able to confine light on this deep-subwavelength scale because they host epsilon-near-zero (ENZ) points near their plasmon frequency, and negative permittivities below this frequency, which produce unique phenomena[11]. However, noble metals are typically chosen for their high-quality plasmons, but they degrade around 300 to 600 °C prohibiting their use in high temperature applications[12]. Additionally, the most commonly used noble metals, Au and Ag, lack CMOS compatibility because they form deep level defects in silicon[13]. Plasmons in metallic transition-metal nitrides (TMNs) have been shown to be both CMOS-compatible and stable at temperatures up to 1,400 °C (under vacuum) at the expense of increased losses[14,15]. Therefore, TMN-based metamaterials can introduce plasmonics to the high temperature regime for applications such as thermophotovoltaics[16], sensing near high temperature processes[17,18], and thermal management[19].

The dispersion relation within a medium relates the momentum and energy of light propagating through the medium, and for an isotropic medium, the relation is $\frac{k^2}{\varepsilon(\omega)} = \frac{\omega^2}{c^2}$ where ε(ω) is the complex permittivity of the medium (ε = $\varepsilon_1$ + i$\varepsilon_2$) as a function of angular frequency, c is the speed of light in a vacuum, k is the wavevector of light, and ω is the angular frequency of light. In E-k space, the isotropic dispersion relation produces a circular constant-energy surface (isofrequency surface) with a finite maximum momentum. For uniaxial anisotropic medium, there are two symmetric axes called the ordinary axes with the same permittivity ($\varepsilon_{ord}$) that we define as the x- and y-directions, and an extraordinary axis in the z-direction with a unique permittivity ($\varepsilon_{ext}$). In this case, the dispersion relation becomes $\frac{k_x^2+k_y^2}{\varepsilon_{ext}} + \frac{k_z^2}{\varepsilon_{ord}} = \frac{\omega^2}{c^2}$ for transverse magnetic (TM) polarized light where the subscript for k denotes the direction of the wavevector. When $\varepsilon_{ext}$ and $\varepsilon_{ord}$ are both positive, the isofrequency surface in E-k space is elliptical with a finite maximum momentum. However, when $\varepsilon_{ext}$ is negative, the isofrequency frequency becomes hyperbolic without a limit to the maximum momentum of light when loss is ignored. However, the maximum momentum becomes finite when loss is accounted for in the system[20,21]. In this case, a signature of hyperbolic dispersion is that the maximum momentum occurs at a propagation angle away from the optical axes. Negative refraction[22], superlensing[23], thermal emission engineering[24] are all observable in hyperbolic media because of the drastic change in how light propagates through it.

Hyperbolic metamaterials are commonly constructed by stacking alternating layers of metals and dielectrics into a superlattice. In this geometry, the in-plane ordinary axes will resemble a metal whose plasmon energy is redshifted from that of the metallic layers while the extraordinary axis will resemble a dielectric. The relative ease of fabrication of this geometry has made it so prolific that it has been used in energy ranges spanning from the deep UV[25] to microwaves[26]. Recently, for high-temperature applications, these superlattices have been made using doped graphene[27,28] or tungsten[29] as the negative permittivity media. However, both of these materials have low energy ENZ points which results in hyperbolicity only occurring at long wavelengths. Alternatively, TMNs possess plasmons in the blue to green region of light enabling hyperbolicity at much shorter wavelengths, including the visible region[30]. TMN-based hyperbolic superlattices have been fabricated using either three-dimensional oxides[31], which increases the probability of oxidation at high temperatures, or three-dimensional nitrides[32–34] as the insulators.

However, the use of two-dimensional hBN as the insulator promises to further increase the thermal stability of hyperbolic superlattices due to its dangling-bond-free nature which prohibits the diffusion of atoms between layers. Additionally, hBN has excellent thermal properties such as its large thermal emissivity[35,36], and a thermal conductivity (≈500 W/mK)[37,38] which is nearly an order of magnitude larger than previously used insulators such as MgO (≈60 W/mK)[39], ScN (8 W/mK)[40], and AlScN (3-8 W/mK)[41].

In this paper, we introduce and study a hyperbolic metamaterial consisting of alternating layers of TiN and two-dimensional hBN. We also fabricate a TiN/$Al_2O_3$ to compare its thermal stability to. The TiN/hBN is found to be unaffected by annealing it at 800 °C for 10 hrs while the TiN/$Al_2O_3$ superlattice is highly degraded. Compared to previous works, our annealing was performed at a relatively high pressure demonstrating the improved stability of our superlattice. The improved thermal stability is due to both the lack of oxygen in the superlattice and the mixed-dimensional interfaces preventing the diffusion of atoms between layers. Additionally, we find that the reduced quality of TiN grown on hBN results in a larger hyperbolic figure of merit (FoM) since the TiN layer is less dispersive to light.

**Results and Discussion**

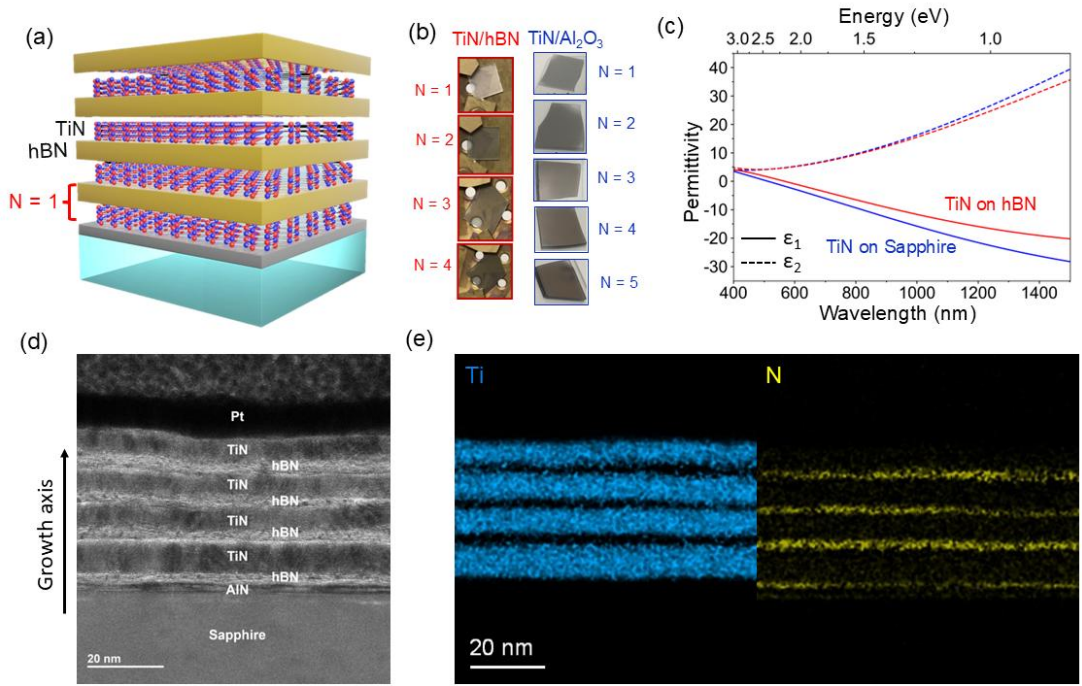

**Figure 1. Assembly and characterization of the TiN/hBN superlattice.** (a) Schematic of the assembled TiN and hBN superlattice (N = 4). (b) Camera images of the superlattice after each TiN deposition for both the TiN/hBN (red) and TiN/$Al_2O_3$ (blue) superlattices. (c) Real (solid lines) and imaginary (dashed) parts of the complex permittivity of TiN (5.03 nm) grown on a sapphire substrate (blue) and TiN (4.57 nm) grown on two-dimensional hBN (red) showing the less negative permittivity characteristic of TiN grown on hBN than sapphire due to lattice mismatch. The refractive index is measured using the VASE ellipsometer. (d) Cross-section TEM and (e) the corresponding EDS maps showing the well-stratified geometry of the TiN/hBN

superlattice. The scale bars in (d) and (e) are 20 nm. Although N appears to only be in the hBN layers, this is the result of the higher crystal quality of hBN yielding a stronger signal.

The stratified nature of the proposed hyperbolic metamaterial is shown in Figure 1a. The metallic TiN layers are deposited by sputtering at 800 °C (See Methods). The process has been shown to produce high-quality films previously[30,42]. The 3 nm thick hBN layers are grown on a sapphire substrate using metal-organic-chemical-vapor-deposition (MOCVD) before they are transferred on top of the TiN using a wet transfer process that has previously produce high-quality, large-area superlattices[43]. For the Al$_2$O$_3$ superlattice, amorphous Al$_2$O$_3$ is grown on TiN using atomic layer deposition (ALD). The robustness of our fabrication process enabled the easy assembly of these superlattices on the centimeter-scale. TiN is absorptive throughout the visible region. As such, both the TiN/hBN and TiN/Al$_2$O$_3$ superlattices appear black, and they become visibly darker upon stacking (Figure 1b).

The quality of the TiN films is studied using standard spectroscopic ellipsometry (See Supporting Information Section S1). By fitting the experimental results to an isotropic Drude-Lorentz model, the complex permittivities and film thicknesses are extracted (Figure S1). The free-carrier contribution to the permittivity ($\varepsilon_{free}$) is given by the equation $\varepsilon_{free}(E) = \frac{-E_p^2}{E^2 + iE\Gamma}$ where $E_p$ and $\Gamma$ are the plasmon energy and damping factor, respectively. The plasmon energy depends on the free-carrier density (N), effective mass ($m_{eff}$), charge of an electron (e), and the vacuum permittivity ($\varepsilon_o$) as $E_p = \hbar\sqrt{\frac{Ne}{\varepsilon_o m_{eff}}}$, and the damping factor is the inverse of mean-scattering time of the free-carriers[44]. The complex permittivities of TiN grown on sapphire and hBN indicate that TiN exhibits better material quality when sapphire is used as the growth substrate compared to hBN (Figure 1c). The improved quality of TiN grown on sapphire is seen as a more negative $\varepsilon_1$, and a redshifted plasmon energy from 2.71 eV on sapphire to 2.65 eV on hBN determined using spectroscopic ellipsometry fitting. Note that the plasmon energies differ from the energy where $\varepsilon_1 = 0$ due to the effects of the background permittivity and intraband resonances. The $\varepsilon_1 = 0$ energies are 2.42 eV and 2.15 eV for TiN grown on sapphire and hBN, respectively. The difference is TiN quality is seen most clearly in the damping factor of the films. The difference in growth quality on the substrates is attributed to lattice mismatch. The lattice constant of TiN is 0.424 nm[45] allowing quasi-epitaxial growth on sapphire[46] whose lattice constant is 0.475 nm (11% mismatch)[47]. However, hBN's lattice constant is 0.25 nm resulting in non-epitaxial growth (41% mismatch)[48]. The decrease in TiN grain size on hBN increases the probability of electron scattering which is reflected in the increased damping factor on hBN of 567 meV compared to 359 meV for TiN grown on sapphire. Although hBN produces lower quality TiN, its dangling-bond-free interface helps with the thermal stability of the superlattice by preventing the diffusion of atoms between layers, and crucially, out-of-plane hyperbolicity is still observed in the metal-insulator superlattice.

The well-stratified geometry is clear in the cross-section transmission electron microscopy (TEM) images (Figure 1d). All of the TiN and hBN layers are clearly resolved from one another. The layered nature of 2D hBN is seen throughout the superlattice showing its high-crystalline quality and dangling-bond-free interface. Additionally, a layer of AlN is observed between the sapphire substrate and the hBN, which may result from nitridation of the sapphire substrate during initial hBN deposition, as the growth substrate is used to fabricate the superlattice on to minimize the number of wet transfer steps that reduce the overall superlattice quality. Between hBN layers, layers of TiN are observed. All of the layers of TiN are of similar brightness in the TEM scan suggesting that they are of similar quality, but the crystal quality is seen to decrease upon stack according to crystallography measurements which we attribute to increased surface roughness (Figure S2). The layered nature of the superlattice is further confirmed by using energy dispersive spectroscopy (EDS) mapping (Figure 1e). Ti is only found in the TiN layers while N appears to only be present in the hBN layers. However, this is the result of the improved crystal quality which yields a stronger N signal in the hBN (Figure S2). In fact, an N signal is observed throughout the superlattice, as confirmed by the log-scale plot in Figure S2c. Similar measurements are performed on the TiN/$Al_2O_3$ (N = 5) superlattice (Figure S3 and S4). The bottom-most TiN shows the best quality within the TiN/$Al_2O_3$ superlattice since it was grown on crystalline sapphire. However, the quality decreased in subsequent TiN layers since they were grown on amorphous, ALD-grown $Al_2O_3$.

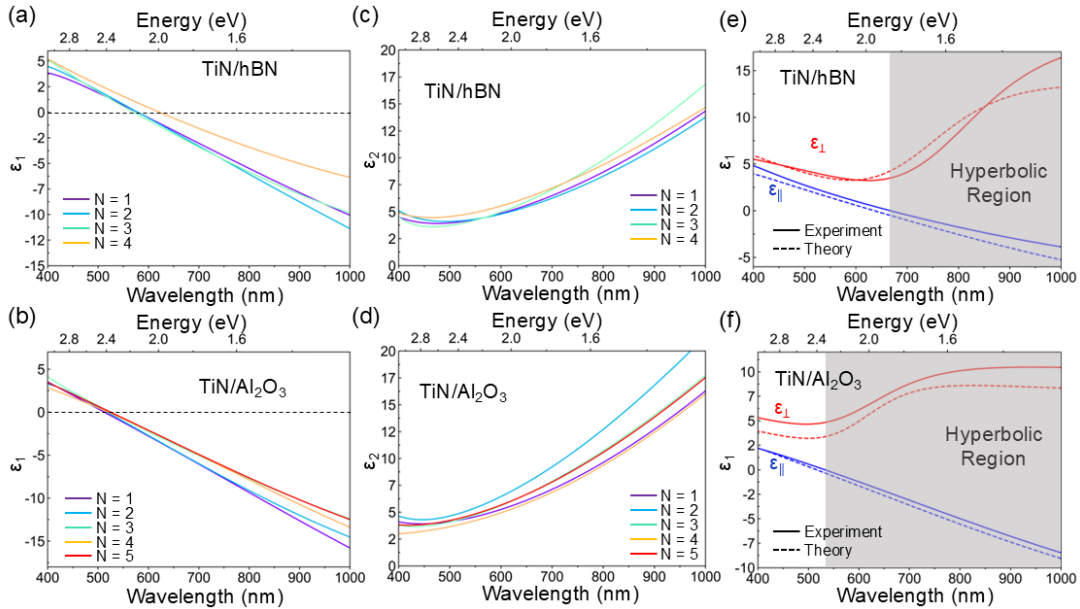

**Figure 2. Permittivities of TiN/hBN and TiN/$Al_2O_3$ superlattices.** The layer-dependent real and imaginary parts of the permittivity of the TiN layers in the **(a, c)** TiN/hBN and **(b, d)** TiN/$Al_2O_3$ superlattices measured using standard spectroscopic ellipsometry. N = 1 denotes the bottom-most TiN layer while N = 4 (N = 5) denotes the top layer for the TiN/hBN (TiN/$Al_2O_3$) superlattice. The theoretical, in-plane (∥) and out-of-plane (⊥) permittivity is calculated using these values for the **(e)** TiN/hBN and **(f)** TiN/$Al_2O_3$ superlattices are compared to the experimentally measured values using Mueller matrix ellipsometry. Ellipsometry was performed using the VASE ellipsometer for figures a-c, the Park Systems ellipsometer for e, and the M-2000 ellipsometer for f.

After the deposition of each TiN layer during fabrication of the TiN/hBN (N = 4) and TiN/Al$_2$O$_3$ (N = 5), spectroscopic ellipsometry is performed to measure the layer-dependent complex permittivity of the TiN films (Figure 2a-d and Figures S5-S6)[49]. The TiN layers are numbered in order of when they were deposited so the N = 1 layers are the bottom-most layers. In the TiN/hBN superlattice, the first three layers are all of similar qualities with $\varepsilon_1$ = 0 wavelengths of 578 nm, 579 nm, and 575 nm from N = 1 to N = 3. However, due to increased surface roughness upon stacking, the quality of TiN layer degrades, and the permittivity is only negative for wavelengths longer than 622 nm. The surface roughness increases the strain within the top TiN film which decreases both the carrier concentration and mobility of TiN[50,51]. This results in the redshifted ENZ point and less-metallic optical properties in the N = 4 layer. Although the effect of strain is not as pronounced in $\varepsilon_2$, the overall plasmonic performance of the fourth layer is reduced in accordance with the common plasmonic figure of merit (FoM = -$\varepsilon_1$/$\varepsilon_2$)[52]. Due to the decreased optical characteristics of the fourth TiN layer, a fifth unit cell was not fabricated as it was for the TiN/Al$_2$O$_3$ superlattice. As for the TiN/Al$_2$O$_3$ superlattice, the variation of permittivity is less than in TiN/hBN superlattice, and this is reflected by the observation that $\varepsilon_1$ = 0 at wavelengths ranging from 513 nm to 525 for all TiN layers. Less strain is accumulated upon stacking within the TiN/Al$_2$O$_3$ superlattice since the ALD process results in a relatively smooth amorphous Al$_2$O$_3$ layer as compared to the wet-transfer process of hBN.

The superlattices are then modeled using an effective medium approximation (EMA) since they consist of deep-subwavelength thick films. According to the EMA, the in-plane effective permittivity ($\varepsilon_\parallel$) is the thickness-weighted average of the permittivities ($\varepsilon_\parallel = \frac{\varepsilon_{TiN} t_{TiN} + \varepsilon_I t_I}{t_{TiN} + t_I}$ where t represents the thickness of the layer and the subscript *I* denotes the insulator), and the out-of-plane effective permittivity ($\varepsilon_\perp$) is the thickness-weighted average of the inverse of the permittivities $\left(\frac{1}{\varepsilon_\perp} = \frac{\frac{t_{TiN}}{\varepsilon_{TiN}} + \frac{t_I}{\varepsilon_I}}{t_{TiN} + t_I}\right)$ [53]. The in-plane effective permittivity therefore resembles a metal since |$\varepsilon_{TiN}$| > |$\varepsilon_I$| at long wavelengths. Additionally, the ENZ point will be redshifted from that of TiN depending on the relative thicknesses of the metal and insulating layers, and the energy of the ENZ point can be predicted using the Drude model (Supporting Information Section S2). The imaginary part of $\varepsilon_\parallel$ will also be less than the metal since the insulator will not contribute to the loss. As for $\varepsilon_\perp$, the real part of the permittivity will remain positive $t_{TiN}$/$\varepsilon_{TiN}$ goes to zero at long wavelengths. Therefore, the effective medium approximation predicts hyperbolic dispersion to occur at long wavelengths depending on both the permittivity and thicknesses of the constitute layers.  The out-of-plane loss will also remain low except for a single peak near the in-plane ENZ energy. Further discussion on the effective medium approximation can be found in the Supporting Information Section S2.

The EMA is measured using Mueller matrix ellipsometry since the Mueller matrix is more sensitive to the out-of-plane permittivity than standard spectroscopic ellipsometry (Supporting Information Section S3)[49]. The EMA permittivity is then extracted from the experimental Mueller matrix by replacing the superlattice with a single layer of the same thickness during modeling (Figures 2e-f and S8-S9). The in-plane permittivity shows excellent agreement between experiment and theory in both superlattices. The ENZ wavelength differs between experiment and theory by 28 nm and 17 nm for the TiN/hBN and TiN/Al$_2$O$_3$, respectively. The out-of-plane permittivity also shows good agreement between experiment and theory, but the agreements is not as strong as the in-plane direction since the signal from the out-of-plane direction is not very strong in the superlattices with cumulative thicknesses of ~ 40 nm. The imaginary part of the EMA permittivity shows similar agreement to the real part (Figure S10). Most importantly, both theory and experiment show hyperbolic regions in the visible and near infrared.

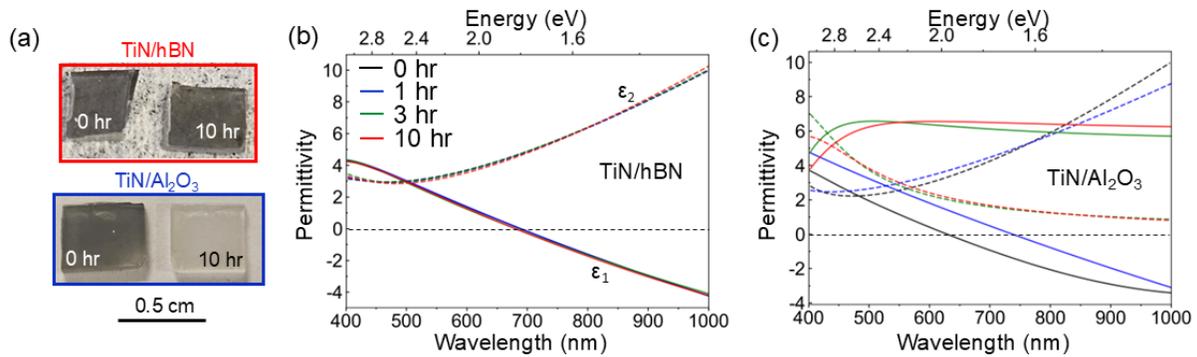

**Figure 3. Effects of annealing on the effective permittivity of the superlattices.** (a) Optical image of the TiN/hBN (red) and TiN/Al$_2$O$_3$ (blue) superlattices before and after 10 hours of annealing in an inert atmosphere at 800 °C. The complex permittivity of the **(b)** TiN/hBN (N = 4) and **(c)** TiN/Al$_2$O$_3$ (N = 5) superlattices after 0, 1, 3, and 10 hours of annealing. The TiN/hBN displayed unchanged optical properties due to the high temperatures while the TiN/Al$_2$O$_3$ superlattice lost its negative permittivity behavior. All of the data in this figure was measured using the Parks System Ellipsometer.

The high-temperature stability of the superlattices is studied by annealing the superlattices at 800 °C for 1 to 10 hours in an Ar environment (See Methods). The annealing is performed at a pressure of 500 mBar which is significantly larger than previous studies on the thermal stability of TMNs that were done at 2x10$^{-6}$ to 10$^{-4}$ mBar (Table S1)[14,30,34,54]. For this measurement, a new TiN/Al$_2$O$_3$ N = 5 superlattice is fabricated with an $\varepsilon_1$ = 0 wavelength closer to that of the TiN/hBN superlattice to allow for a more direct comparison between the two. The TiN/hBN is observed to have maintained it grey, absorptive appearance after 10 hours of annealing as shown in Figure 3a. However, the TiN/Al$_2$O$_3$ superlattice becomes transparent upon annealing suggesting that there is no longer any absorptive, plasmonic media remaining. The continued hyperbolic properties of the TiN/hBN superlattice is confirmed by spectroscopic ellipsometry (Figure 3b and Figure S11). The $\varepsilon_1$ = 0 wavelength redshifts by only 6 nm after annealing for 10 hours. The thermal stability of the TiN/hBN is attributed both to the lack of oxygen within the superlattice and the mixed-dimensional nature of the layers. The mixed-dimensional nature results of the TiN/hBN being resistant to diffusion of atoms across the interface, and therefore, the stratified geometry remains even after exposure to high temperatures.

The lack of oxygen in the dielectric means that there are not any oxygen for the Ti atoms to bond to that would degrade the TiN which has also been studied in previous all-nitrogen geometries[34]. Although the annealing is done in an inert environment, in practice, the superlattice can be encapsulated by a thick layer of hBN to protect it from oxygen and moisture[55]. This would enable the use of the TiN/hBN hyperbolic superlattice in oxygen-rich environments that reach high temperatures.

The presence of oxygen in the dielectric caused the plasmonic nature of the TiN/Al$_2$O$_3$ superlattice to disappear upon annealing (Figure 3c). After 1 hr of annealing, the superlattice last still has a negative effective permittivity at long wavelengths, but the $\varepsilon_1 = 0$ wavelength has redshifted by 109 nm (286 meV). Further annealing then eliminates the negative permittivity behavior, and the stratified, anisotropic optical properties are no longer observed (Figure S12). This is because oxygen more easily bonds to Ti than Al atoms[56,57]. Therefore, at high temperatures, the oxygen can diffuse across the all-3D interface and oxidize the TiN. The TiN is then replaced with insulating TiO$_2$ which is seen as a positive real part of the permittivity at all wavelengths, and a reduction imaginary part of the permittivity. Full conversion to TiO$_2$ would result in $\varepsilon_2 = 0$ at all wavelengths studied. However, the reduction in $\varepsilon_2$ from 3 hrs of annealing to 10 hrs indicates that the conversion process is incomplete, and that there is still some absorptive TiN remaining.

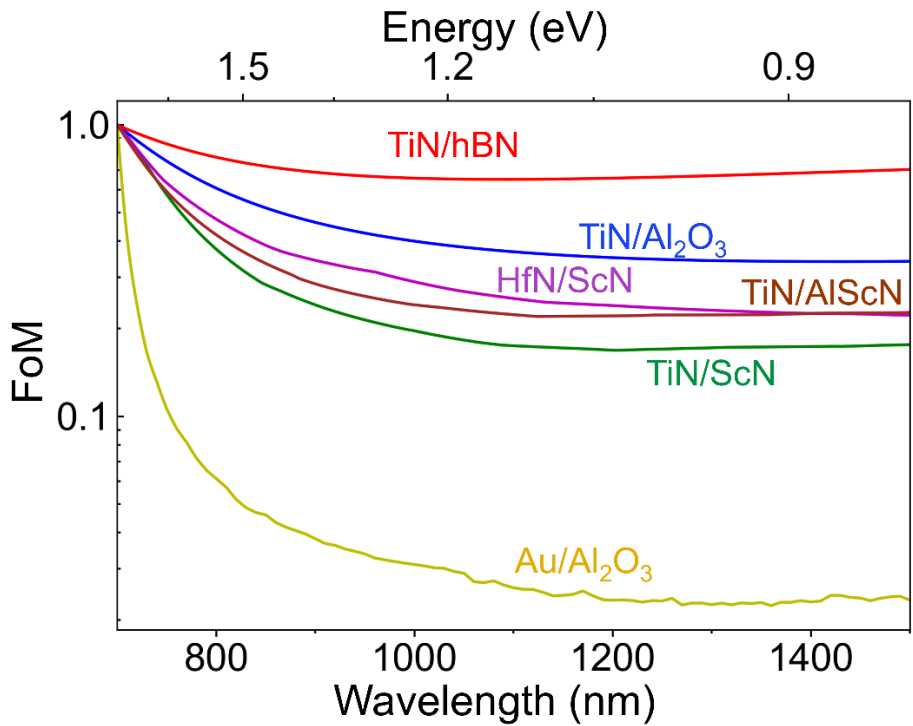

**Figure 4. Figure of merit of different superlattices.** The figure of merit (FoM = Re($\beta_\perp$)/Im($\beta_\perp$)) of several different hyperbolic superlattice geometries which are commonly found in literature for high-temperature applications[32–34] along with an Au/Al$_2$O$_3$ superlattice for reference. The fill fractions of the metals and dielectrics have been normalized such that all of the superlattices are hyperbolic for wavelengths > 700 nm to improve the comparison.

The commonly accepted FoM for a hyperbolic superlattice is $Re(\beta_\perp)/Im(\beta_\perp)$ where $\beta_\perp$ is the complex propagation of constant of light moving perpendicular to the layers (parallel to the normal)[32]. As the propagation constant can be written in terms of the complex refractive index ($\beta = 2\pi(n + ik)/\lambda$ where n and k are the real and imaginary parts of the complex refractive index, respectively), the FoM can be rewritten as the ratio of the in-plane effective refractive index ($n_{eff}$) and the in-plane effective extinction coefficient ($k_{eff}$). Therefore, the ideal hyperbolic structure allows for a large degree of phase accumulation ($n_{eff} \gg 1$) and a low amount of loss ($k_{eff} \ll 1$). However, hyperbolicity requires that $k_{eff} > n_{eff}$, or else the effective permittivity would be positive ($\varepsilon_1 = n^2 - k^2$). This means that the upper limit of the FoM is 1, and this limit occurs when $\varepsilon_1 = 0$ in-plane.

In Figure 4, the FoM of several hyperbolic superlattice geometries are compared. The geometries include Au/Al$_2$O$_3$, to show the performance of a classical, low-temperature TMN superlattices[32,33], previously studied superlattices for high-temperature applications under high vacuum[34] and our superlattices. Since the thickness of each layer can be easily controlled during fabrication, we have chosen to vary the relative thicknesses of the metals and dielectric such that the hyperbolic region begins at 700 nm for easy comparison between media. Of all the superlattices studied, the most important value for a high FoM is the dispersion of the metallic layer. The lower quality TiN on hBN shows the largest FoM throughout the hyperbolic region, followed by other TMN superlattices, and then Au/Al$_2$O$_3$ has the lowest FoM. Therefore, it is observed that less dispersive metals are preferred at energies below the hyperbolic transition since they result in less negative permittivities (relatively large $n_{eff}$ compared to $k_{eff}$). Although the decreased group velocity of TiN/hBN is beneficial for the FoM when the superlattices are normalized such that hyperbolicity occurs at $\lambda = 700$ nm, the drawback is that the permittivity does not become negative until longer wavelengths as shown in Figure 1c. Therefore, it cannot be used to produce a hyperbolic superlattice at wavelengths as short as quasi-epitaxial TiN would enable. Additionally, the lower crystal quality of TiN would suggest that it is less stable than quasi-epitaxial TiN, but we have demonstrated that the lower quality TiN/hBN superlattice is still stable after being heated to 800 °C for 10 hours.

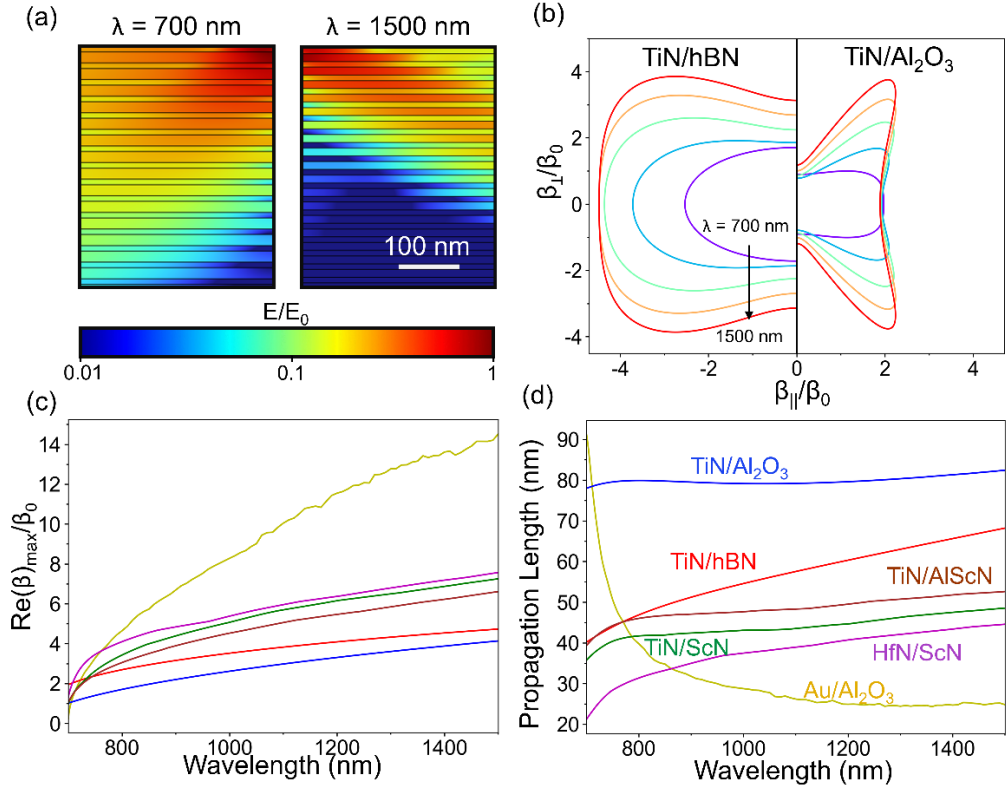

**Figure 5. Dispersion of hyperbolic superlattices. (a)** The simulated electric field distribution (on a logarithmic scale) within the TiN/hBN superlattice at λ = 700 nm (left) and 1500 nm (right) normalized to the incident electric field showing the improved propagation at longer wavelengths without increased loss. The system is simulated using COMSOL Multiphysics, and the top-most layer is TiN followed by hBN. **(b)** Dispersion of the TiN/hBN (left) and TiN/Al$_2$O$_3$ (right) superlattices at wavelengths varying from 700 nm to 1500 nm at 200 nm increments. The wavelength-dependent **(c)** maximum normalized real part of the wavevector and **(d)** propagation length along the direction of maximum real part of the wavevector for different hyperbolic superlattices whose geometry is chosen such that hyperbolicity begins at λ = 700 nm.

To further investigate the hyperbolic dispersion within the superlattices, the electric field distribution for light within the TiN/hBN is simulated using COMSOL Multiphysics (Figure 5a). The light is excited using a point source placed 25 nm above the superlattice. A point source is used since the incident angle of light will therefore vary laterally such the propagation angle can be easily observed. By comparing the electric field at λ = 700 nm and 1500 nm, it is observed that the longer wavelength results in more directional propagation within the superlattice which is a result of the more negative ε$_1$. At λ = 700 nm, the propagation resembles that of an elliptical system which we confirm calculating the dispersion relation between β$_∥$ and β$_⊥$ while accounting for loss in the system (Figure 5b)[21]. The loss in the system causes the dispersion to remain elliptical shape when ε$_1$ ≲ 0 since the loss limits the maximum possible momentum of light in the medium[20,21]. In this case, the direction of maximum momentum remains near one of the optical axes. However, as the wavelength increases, and ε$_{1,∥}$ becomes more negative, the dispersion bulges and the maximum momentum occurs at a direction away from the optical axes. This transition is most clearly seen in the dispersion of the TiN/Al$_2$O$_3$ and the other superlattices (Figure S13). The superlattices in Figure 4, whose FoM were compared, are further compared to one another using their wavelength-dependent maximum momentum and propagation length along that direction (Figures 5c and 5d). All of the superlattice show relatively low momenta values near the hyperbolic transition, but at longer wavelengths, Au/Al$_2$O$_3$ shows the biggest enhancement, followed by the quasi-epitaxial TMN on AlScN or ScN superlattices from literature, and lastly, our superlattices. In this, there is a clear trend that the more negative the in-plane permittivity is, the larger the maximum momentum is. This observation is consistent with previous theory that predicts that highly negative permittivities and low loss functions ($\delta = \varepsilon_2/\varepsilon_1$) yields larger momenta $(\mathrm{Re}(\beta))_{max} \propto \sqrt{(|\varepsilon_\parallel| + |\varepsilon_\perp|)/(|\delta_\parallel| + |\delta_\perp|)}$ )[21]. However, we observe that the trend reverses for the propagation length in these media. The lower quality TiN superlattices host longer propagation lengths than the quasi-epitaxially grown TMN layers. Therefore, there is a necessary trade-off between maximum momentum and propagation length in these superlattices.

**Conclusion**

We have fabricated both TiN/hBN and TiN/Al$_2$O$_3$ superlattices to explore the effects of mixed-dimensional interfaces and oxide layers on the thermal stability of their stratified nature. Their stratified nature results in hyperbolic effective permittivities. The hyperbolicity of the TiN/hBN is unaffected upon annealing at 800 °C for 10 hrs. The stability is the result of the lack of oxygen in the superlattice and the mixed-dimensional superlattice preventing the diffusion of atoms between layers. Importantly, we found the hyperbolic superlattice to be stable at high temperatures at higher pressures than previous works on all-3D hyperbolic superlattices. However, the TiN/Al$_2$O$_3$ is found to completely degrade during the annealing process. Although the quality of TiN is reduced when grown on hBN, instead of quasi-epitaxially grown on sapphire, the FoM of the hyperbolic superlattice is found to increase due to the reduced lossy and dispersive nature of the lower quality TiN layers. Additionally, we observed that the non-epitaxial TMN superlattices supported hyperbolic waves with much larger propagation lengths than quasi-epitaxial ones. Hyperbolic superlattices are vital for optical thermal management, sub-wavelength focusing, and cloaking, and our work demonstrates that TiN/hBN superlattices can be used to enable these technologies in high-temperature systems.

**Methods**

*Superlattice Fabrication.*

TiN/Al$_2$O$_3$ superlattice

A quasi-epitaxial TiN layer was grown on c-plane sapphire using radio frequency magnetron sputtering at 800 °C to achieve good metallic properties. Subsequently, a 3 nm-thick Al$_2$O$_3$ layer was deposited *via* ALD using a custom-designed system developed by the local company Syskey Technology. Finally, the alternating growth of TiN and Al$_2$O$_3$ was repeated to form the superlattice (N = 5). The detailed growth parameters are provided in Supporting Information Section S4.

TiN/hBN superlattice

A quasi-epitaxial TiN layer was grown on MOCVD-grown hBN thin films using radio frequency magnetron sputtering at 800 °C to achieve good metallic properties. Next, Wet chemical transfer was used to move the MOCVD-grown hBN thin films from its parent sapphire substrate to the superlattice. Approximately 200 nm poly(methyl methacrylate) (PMMA) 950kA4 was spin coated on 1 × 1 $cm^2$ hBN samples and kept in air to dry overnight. The PMMA coated samples were dipped in deionized (DI) water and heated to 85 °C on a hot plate for 1 hr until air bubbles were observed. Then, samples were slowly dipped at a 45° angle in 3 M potassium hydroxide (KOH) solution at 85 °C such that the hBN−PMMA delaminates from the substrate and floats on top of the solution. Using a clean glass slide, the floating PMMA-supported hBN film was transferred to fresh DI water, and this step was repeated multiple times to remove any residual contamination from the delamination process. Finally, the floating PMMA-supported hBN film was scooped onto the superlattice. After allowing the sample to dry for 6 hours, it was placed on a hot plate at 70 °C for 4 hours to improve adhesion to the superlattice. The PMMA was removed by immersion in acetone for 1 hour at 45 °C. Lastly, the process of TiN growth and hBN transfer was repeated to complete the superlattice (N = 4). The detailed growth parameters are provided in Supporting Information Section S4.

*Spectroscopic Ellipsometry*

Three different ellipsometers were used in this work. As such, we have labeled which ellipsometer was used in the figure description where spectroscopic ellipsometry data is presented. The first ellipsometer is VASE with high accuracy, which is used to examine the optical quality of deposited thick films. The spectral range is from 300 to 2000 nm. Another ellipsometer was the RC-2 Park Systems imaging micro-ellipsometer. This ellipsometer has a resolution of ~1 µm, and it can perform Mueller matrix ellipsometry from 400 to 1,000 nm. Lastly, the third ellipsometer used is the J.A. Woollam M-2000 ellipsometer with a focusing lens that made the spot size ~ 100 µm, and it has a wavelength range of 371 to 1687 nm.

*TEM*

Samples were prepared using the pre-thinning method. The specimens were thinned to approximately 0.1 µm using a focused ion beam (FIB, FEI Helios) and subsequently characterized by TEM and EDS-mapping performed with the FEI Talos F200X.

*Annealing*

Annealing was performed using an Lindberg Blue M Box Furnace from Thermo Scientific at 800 °C at pressures are 500 mTorr with 50 standard cubic centimeter per minute (sccm) of Ar flowing through furnace through the entire process.

**Acknowledgements**

D. J. and J. L. acknowledge primary support for this work from the Asian Office of Aerospace Research and Development (AOARD) of the Air Force Office of Scientific Research (AFOSR) FA2386-21-1-4063. D.J. and B. C. acknowledge partial support from the Office of Naval Research (ONR) Young Investigator Award (YIP) (N00014-23-1-203) Metamaterials Program. D.J. acknowledges partial support from the Alfred P. Sloan Foundation's Sloan Fellowship in Chemistry. Y.J.L. acknowledge financial support from National Science and Technology Council, Taiwan (Grant No. NSTC-110-2124-M-001-008-MY3, NSTC-113-2112-M-001-014), and Academia Sinica (AS-TP-113-M02). A portion of the sample fabrication, assembly, and characterization were carried out at the Singh Center for Nanotechnology at the University of Pennsylvania, which is supported by the National Science Foundation (NSF) National Nanotechnology Coordinated Infrastructure Program Grant NNCI-1542153.

**Supporting Information:**

**High-Temperature Resilient Hyperbolicity in a Mixed-Dimensional Superlattice**

Jason Lynch[1, †], Tzu-Yu Peng[2, 3, †], Jing-Wei Yang[2, 3], Ben R. Conran[4], Bongjun Choi[1], Cindy Yueli Chen[5], Zahra Fakhraai[5], Clifford McAleese[4], Yu-Jung Lu[2, 3, *], Deep Jariwala[1, *]

[1]Department of Electrical and Systems Engineering, University of Pennsylvania, Philadelphia, Pennsylvania 19104, United States

[2]Research Center for Applied Sciences, Academia Sinica, Taipei, Taiwan

[3] Graduate Institute of Applied Physics, National Taiwan University, Taipei, Taiwan

[4]AIXTRON Ltd, Swavesey, Cambridge CB24 4FQ, UK

[5]Department of Chemistry, University of Pennsylvania, Philadelphia, PA 19104, USA

[*]Corresponding Authors: yujunglu@gate.sinica.edu.tw , dmj@seas.upenn.edu

[†]These authors contributed equally to this work

## Section S1. Standard Spectroscopic Ellipsometry

Standard spectroscopic ellipsometry measures the difference in complex reflection coefficient ($\rho = \frac{r_{TM}}{r_{TE}} = \tan(\psi)\, e^{i\Delta}$) for transverse electric (TE) and transverse magnetic (TM), and using a model to recreate these differences, the isotropic permittivity can be accurately calculated[49]. The metallic contribution to permittivity is modelled by the Drude-Lorentz model:

$$\varepsilon_{Drude}(E) = \frac{-E_p^2}{E^2 + iE\Gamma}$$

where E is the energy of incident light, $E_P$ is the plasmon energy, and $\Gamma$ is the damping factor. $E_p$ and $\Gamma$ are then treated as fit parameters. The contribution of the $i^{th}$ intraband transitions are then modelled using the Lorentz model:

$$\varepsilon_{Lorentz}(E) = \frac{f_i E_i \Gamma_i}{E_i^2 - E^2 - iE\Gamma_i}$$

where $f_i$, $E_i$, and $\Gamma_i$ are the oscillator strength, resonant energy, and the damping factor of the $i^{th}$ transition, respectively, and all three are fit parameters. The total permittivity is then the sum of the Drude-Lorentz model, a series of Lorentz models, and a background permittivity ($\varepsilon_\infty$):

$$\varepsilon(E) = \varepsilon_\infty + \varepsilon_{Drude} + \sum_i \varepsilon_i^{Lorentz}(E)$$

The fit parameters are then varied to minimize the root-mean-square-error (RMSE):

$$RMSE = \sqrt{\frac{1}{2p-q} \sum_i \left( \left(\psi_i^{mod} - \psi_i^{exp}\right)^2 + \left(\Delta_i^{mod} - \Delta_i^{exp}\right)^2 \right)}$$

where p and q refer to the number of wavelengths measured and the number of fit parameters, respectively, the sum is over the wavelengths, and the superscripts "mod" and "exp" denote the modeled and experimental values, respectively. Typically, an RMSE below 10 indicates an accurate model. The modeling was done using the CompleteEase software from J. A. Woollam.

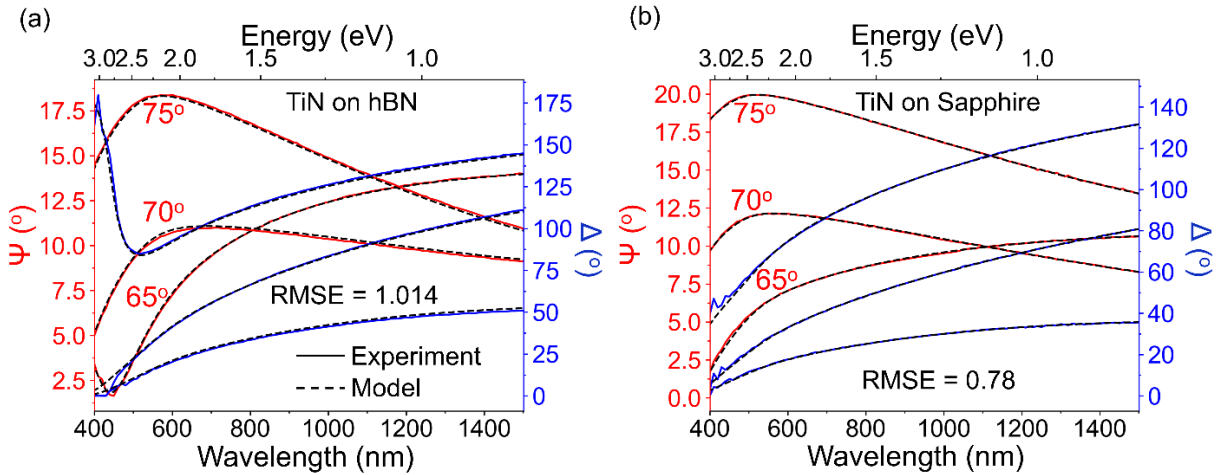

**Figure S1. Standard spectroscopic ellipsometry for TiN grown on hBN and Sapphire.** Experimental (solid lines) and model (dashed) Ψ and Δ at AOI's of 65°, 70°, and 75° for TiN grown on **(a)** hBN and **(b)** sapphire. The calculated complex permittivity can be seen in Figure 1b, and a discussion on the fitting procedure is given below. The measurements were performed using the VASE ellipsometer.

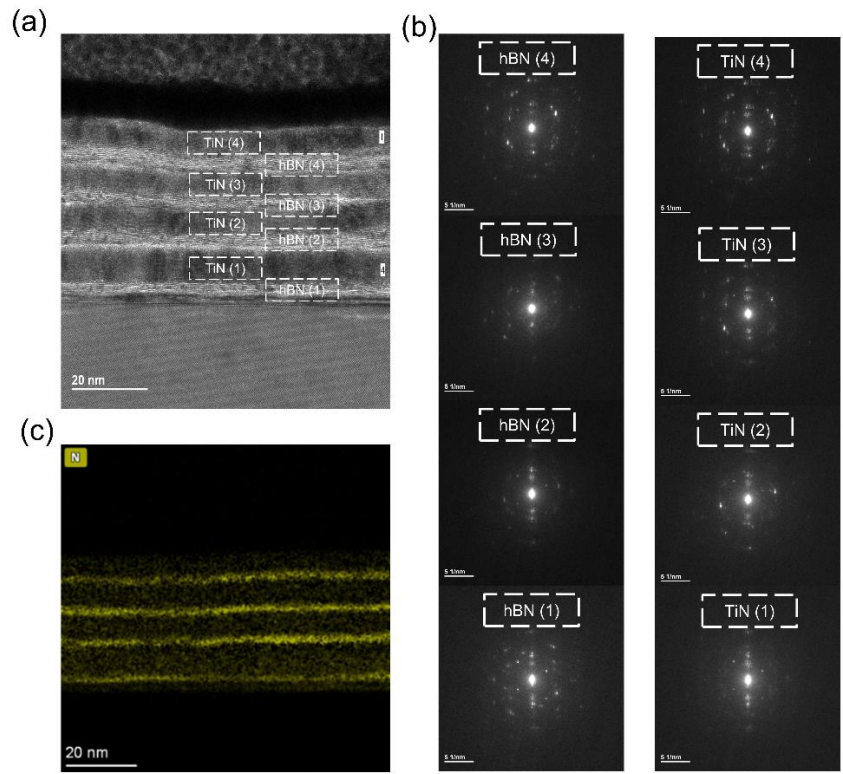

**Figure S2. Crystallography of TiN/hBN (N = 4) superlattice. (a)** TEM image of the superlattice with each layer labelled for the crystallography. The layers are numbered in terms of when the layer was deposited, (1) at the bottom and (4) at the top. **(b)** The crystallography of the hBN (left) and TiN (right) layers with scale bars of 5.1 nm$^{-1}$. **(c)** Atomic mapping of N on a log-scale showing a nitrogen signal in both the hBN and TiN layers. The spot size for TEM/EDS measurements was approximately 2 nm, while the beam size for nano-beam diffraction (NBD) was around 5 nm.

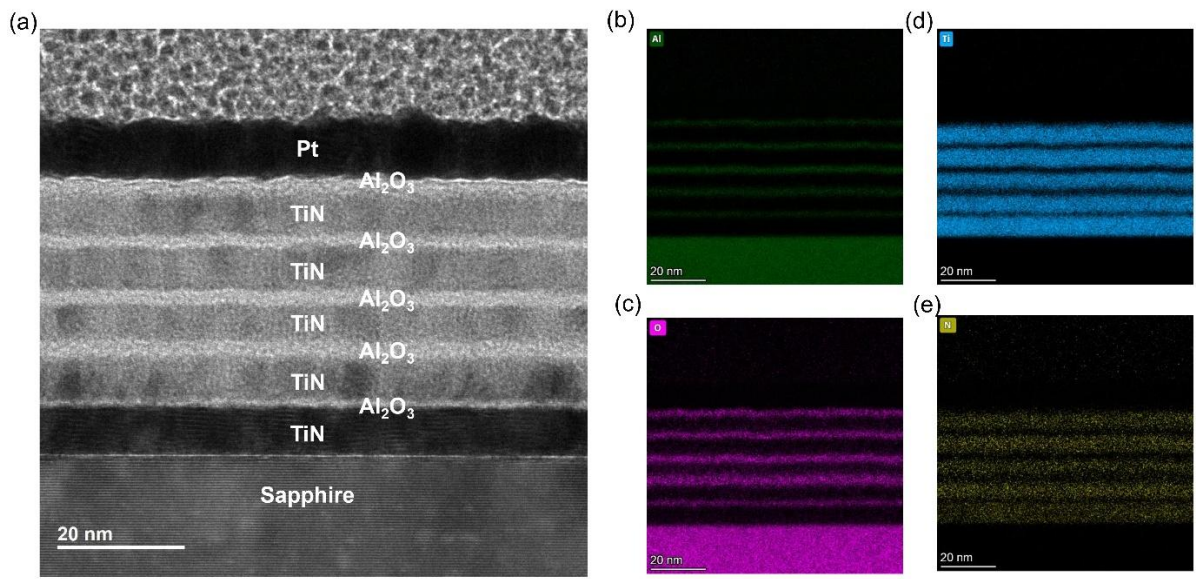

**Figure S3. TEM of the TiN/Al₂O₃ superlattice (N = 5). (a)** TEM scan of the TiN/Al₂O₃ superlattice (N = 5), and atomic mapping of **(b)** Al, **(c)** O, **(d)** Ti, and **(e)** N. All scale bars are 20 nm. The regions of Al and O overlap one another.

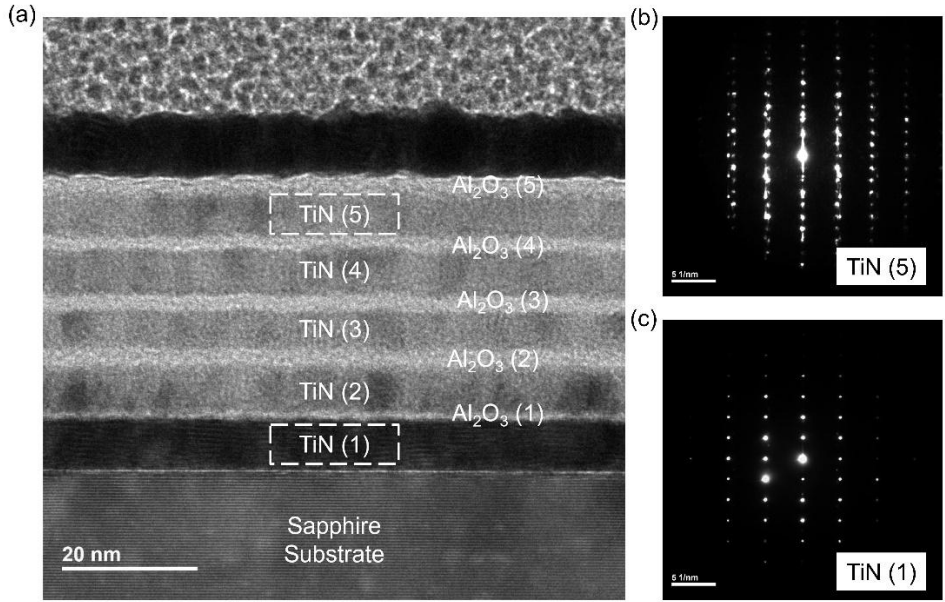

**Figure S4. Crystallography of the TiN/Al₂O₃ (N = 5) superlattice. (a)** TEM scan of the superlattice. The bottom-most layer appears visibly darker than the other TiN layers because of its improved crystallinity as it is epitaxially grown on sapphire while the others are grown on amorphous, ALD-grown Al₂O₃ layers. This is confirmed using crystallography of the **(b)** top-most and **(c)** bottom-most TiN layers which have scale bars of 5.1 nm$^{-1}$.

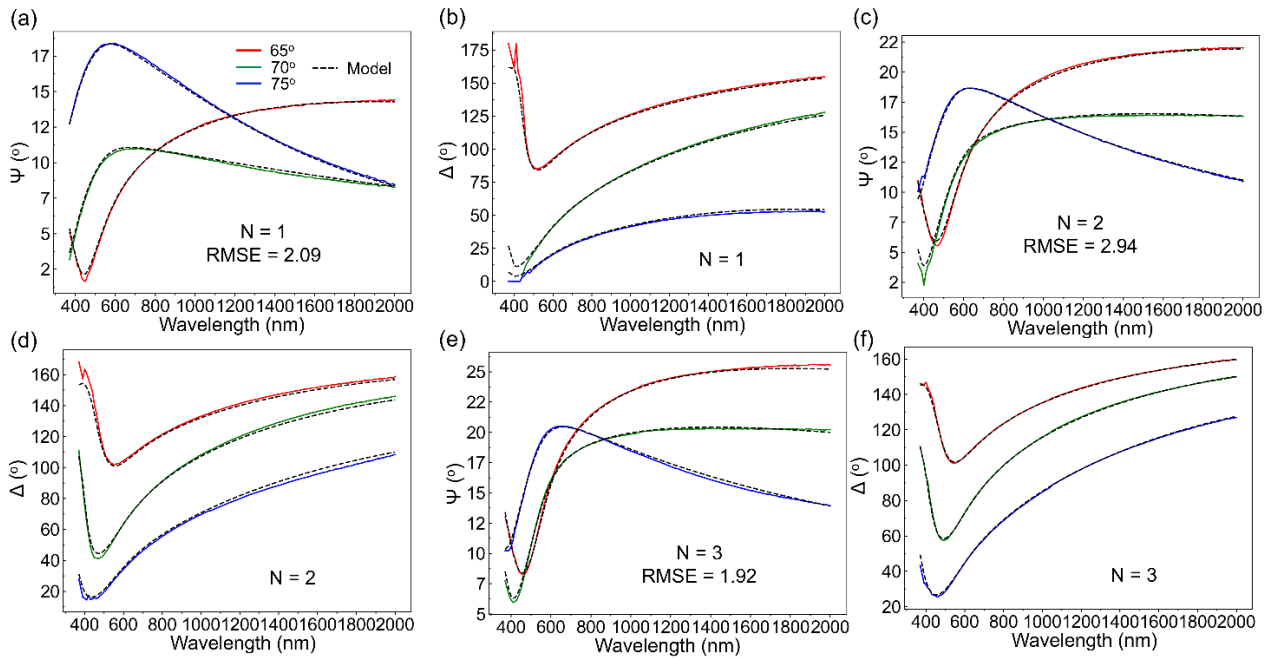

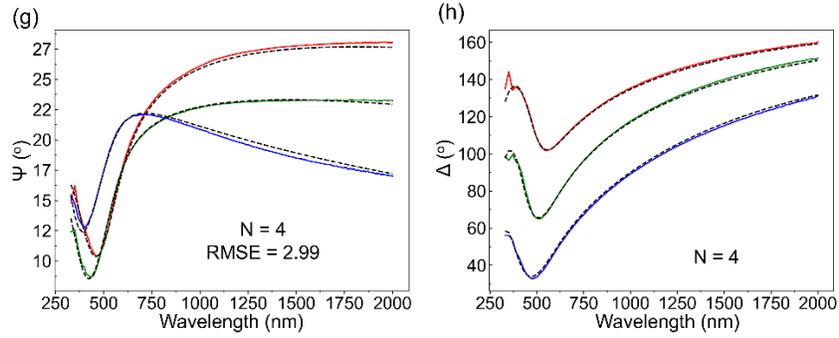

**Figure S5. Layer-dependent ellipsometry of the TiN/hBN (N = 4) superlattice.** Ψ and Δ for the TiN/hBN superlattice during fabrication at incident angles of 65° (red), 70° (green), and 75° (blue). Ellipsometry was performed using the VASE ellipsometer after the deposition of each TiN layer, and the layer-dependent permittivity is extracted using the model and found in Figure 2a-b. The RMSE uses the equation in section S1, and the RMSE is shown on the graphs for Ψ despite both Ψ and Δ being used to calculate it.

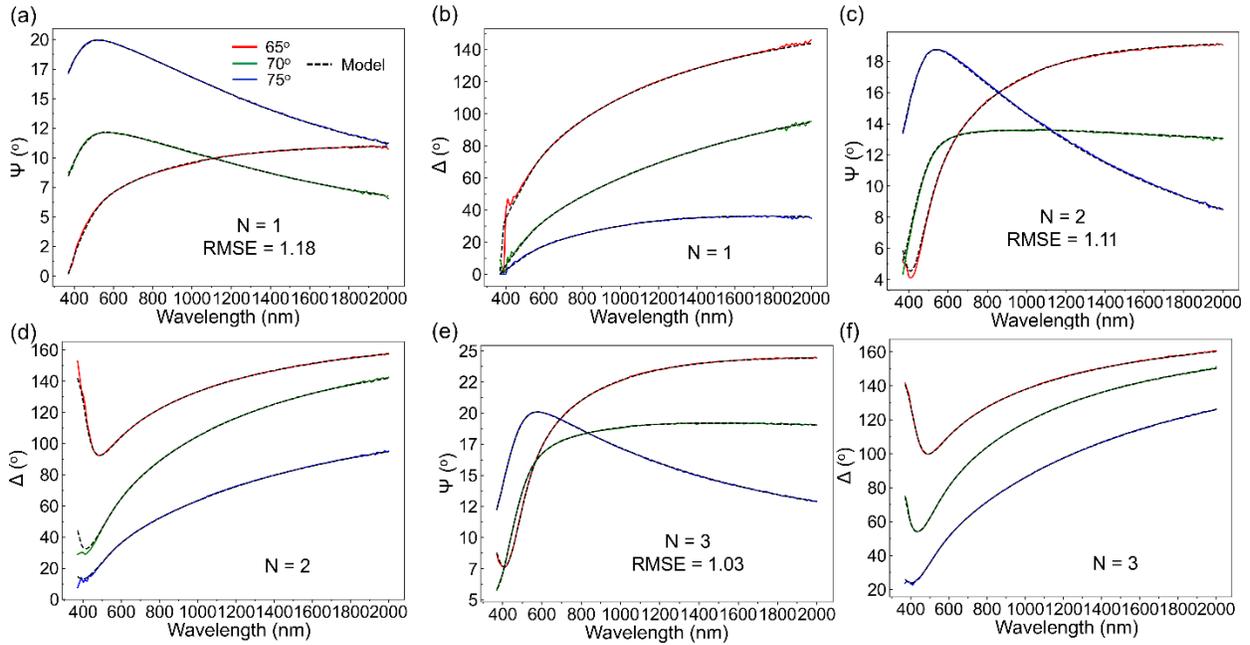

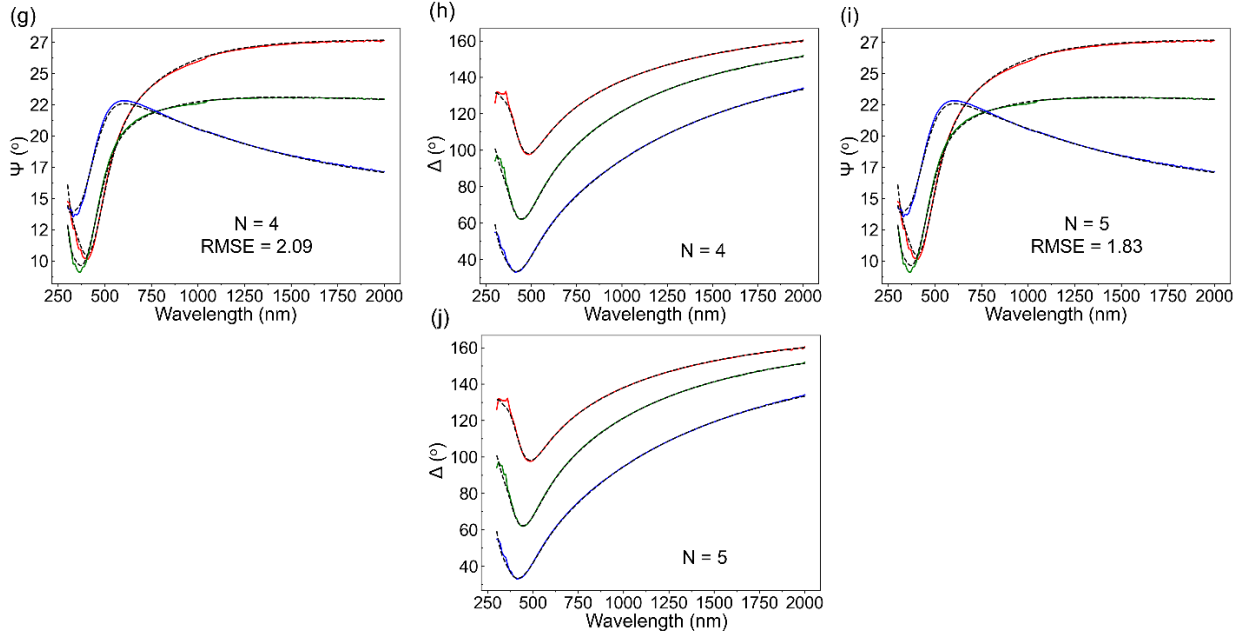

**Figure S6. Layer-dependent ellipsometry of the TiN/Al$_2$O$_3$ (N = 5) superlattice.** Ψ and Δ for the TiN/Al$_2$O$_3$ superlattice during fabrication at incident angles of 65° (red), 70° (green), and 75° (blue). Ellipsometry was performed using the VASE ellipsometer after the deposition of each TiN layer, and the layer-dependent permittivity is extracted using the model and found in Figure 2c-d. The RMSE uses the equation in section S1, and the RMSE is shown on the graphs for Ψ despite both Ψ and Δ being used to calculate it.

## Section S2. Effective Medium Approximation

When a superlattice is made of layers whose thicknesses are much less than the wavelength of light within the media, an effective medium approximation (EMA) can be used to model the entire superlattice as a single medium[53]. According to the EMA, the effective in-plane permittivity is $\varepsilon_{\parallel} = \frac{\varepsilon_{TiN} t_{TiN} + \varepsilon_I t_I}{t_{TiN} + t_I}$, where t represents the thickness of the layer and the subscript *I* denotes the insulator, and the effective out-of-plane permittivity is $\frac{1}{\varepsilon_{\perp}} = \frac{\frac{t_{TiN}}{\varepsilon_{TiN}} + \frac{t_I}{\varepsilon_I}}{t_{TiN} + t_I}$. As discussed in the main manuscript of this work, the out-of-plane permittivity will be positive for all wavelengths studied, and the in-plane permittivity will be negative at wavelengths longer than the epsilon-near-zero (ENZ) point of the bulk plasmonic medium. When the effective permittivities have different signs, the superlattice is in the hyperbolic regime. The wavelength where this occurs depends on both the permittivities and relative thicknesses of the constituent layers:

$$\varepsilon_{\parallel} < 0 \text{ when } |Re(\varepsilon_{TiN})| \geq \varepsilon_I * \frac{t_I}{t_{TiN}}$$

Therefore, we can see that using a metal with more negative permittivity, a lower permittivity insulator, or increasing the plasmonic layer thickness relative to the insulating layer will all cause hyperbolicity to occur at shorter wavelengths.

We can further investigate the EMA by modeling the plasmonic permittivity as the sum of the background permittivity and the Drude-Lorentz model ($\varepsilon_{TiN} = \varepsilon_\infty - \frac{E_p^2}{E^2+iE\Gamma}$, see Section S1). The contributions of the intraband transitions are currently being ignored for both simplicity and because the resonances do not contribute significantly to the permittivity at the energies where the EMA becomes negative in our superlattices. The real part of the plasmonic permittivity is then:

$$Re(\varepsilon_{TiN}) = \varepsilon_\infty - \frac{E_p^2}{E^2 + \Gamma^2}$$

By plugging this equation into the EMA and solving for energy, the energy of the elliptical-to-hyperbolic transition is calculated as

$$E = \sqrt{\frac{E_p^2}{\varepsilon_\infty - \varepsilon_I * \frac{t_I}{t_{TiN}}} - \Gamma^2}$$

Here, we can see that hyperbolicity can occur at higher energies as $E_p$ ($\varepsilon_\infty$, $\Gamma$) increases (decreases).

## Section S3. Mueller Matrix Ellipsometry

In isotropic media, the optical constants at each wavelength is described by two parameters (the real and imaginary part of the permittivity). Due to this, standard spectroscopic ellipsometer (See Section S1) can accurately determine the complex permittivity since it also measures two parameters ($\Psi$ and $\Delta$, See Section S1). However, in anisotropic media, the real and imaginary part of the permittivity must be determined for each optical axis. For uniaxial media, such as the layered superlattice in this work, the two perpendicular in-plane directions have the same permittivity ($\varepsilon_\parallel$), and the out-of-plane has its own unique permittivity ($\varepsilon_\perp$). Therefore, not enough information is collected by a single standard spectroscopic ellipsometry measurement to accurately determine the optical constants of an anisotropic media.

Although out-of-plane anisotropy can be probed using multiple angles of incidence, and in-plane anisotropy can be probed by rotating the sample in-plane, the simplest way to determine the anisotropic permittivity of a media is Mueller matrix ellipsometry[49]. The 4 x 4 Mueller matrix relates the Stokes vector of incident and reflected to fully capture the three-dimensional light-matter interactions. Additionally, the Mueller matrix is normalized such that $M_{11}$ = 1 by convention which results in 15 independent values being measured at each wavelength.

The anisotropic permittivity is then calculated the same way as an isotropic medium as described in section S1 with two main differences. First, each optical axis will have its own background permittivity, Drude-Lorentz model, and series of Lorentz oscillators. Second, the RMSE is now calculated by the equation

$$RMSE = \sqrt{\frac{1}{15p - q} \sum_{i,j} \left(M_{ij}^{mod} - M_{ij}^{exp}\right)^2}$$

Where the sum is over all of the 15 independent elements of the Mueller matrix.

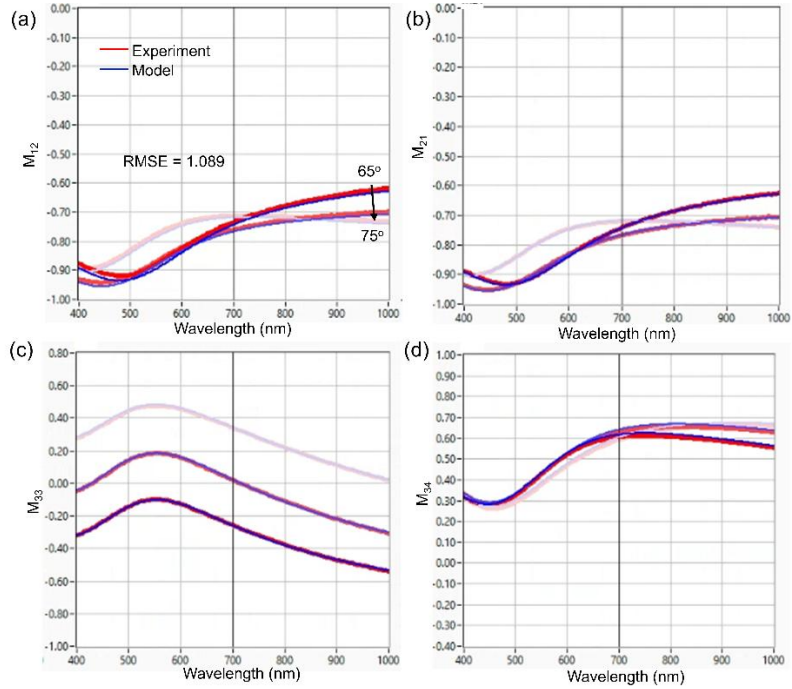

**Figure S8. Select Mueller matrix elements for the pre-annealing TiN/hBN superlattice.** The experimental and model values for **(a)** $M_{12}$, **(b)** $M_{21}$, **(c)** $M_{33}$, and **(d)** $M_{34}$ for the TiN/hBN superlattice using the Park Systems Ellipsometer. The experiment and model show good agreement (RMSE = 1.089). The modeled uniaxial permittivity can be found in Figure 2e and Figure S10.

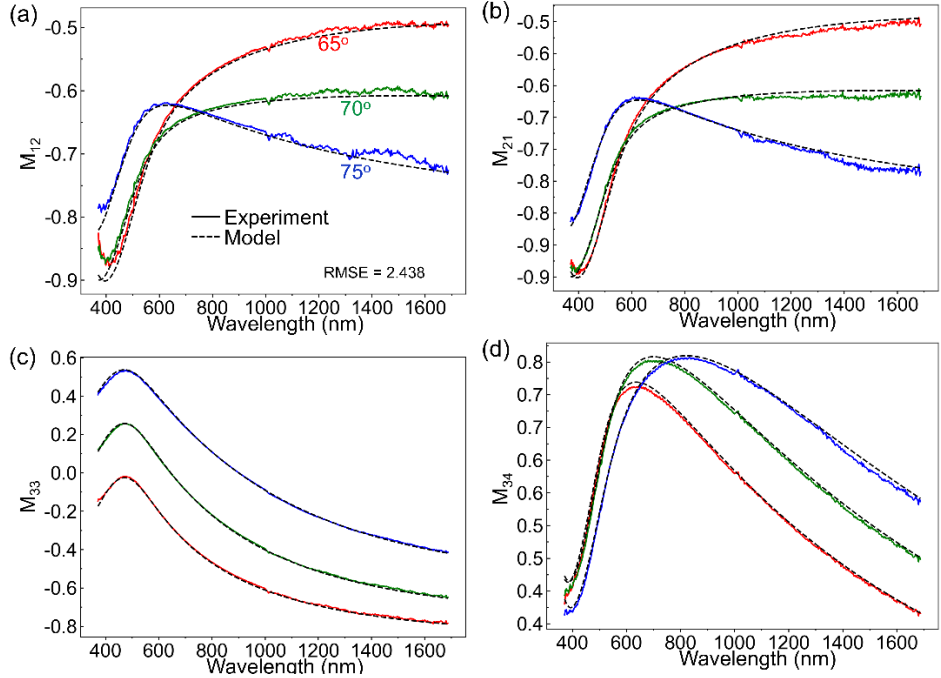

**Figure S9. Select Mueller matrix elements for the pre-annealing TiN/Al$_2$O$_3$ superlattice.** The experimental and model values for **(a)** M$_{12}$, **(b)** M$_{21}$, **(c)** M$_{33}$, and **(d)** M$_{34}$ for the TiN/Al$_2$O$_3$ superlattice using the M-2000 Ellipsometer. The experiment and model show good agreement (RMSE = 2.438). The modeled uniaxial permittivity can be found in Figure 2f and Figure S10.

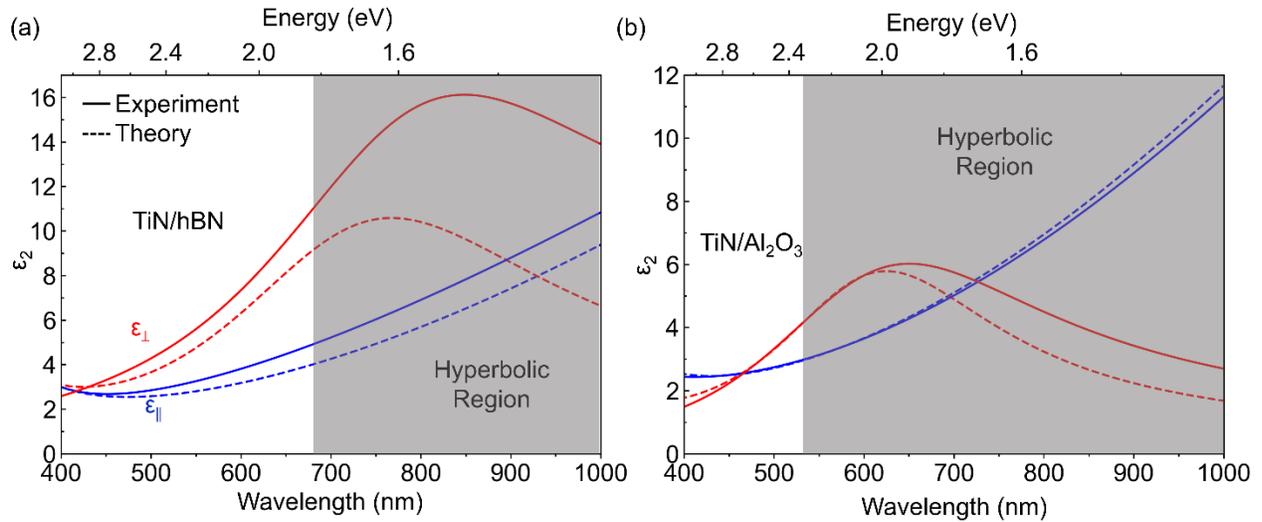

**Figure S10. Imaginary part of the permittivity of the pre-annealing EMA.** Imaginary part of the EMA for the **(a)** TiN/hBN and **(b)** TiN/Al$_2$O$_3$ superlattices. The real part of the permittivities are shown in Figures 2e-f.

**Table S1. Annealing conditions of transition metal nitride (TMN) films in literature.**

| Material | Vacuum degree (mbar) | Annealing temperature(°C) | Annealing time |
|---|---|---|---|
| TiN Film[14] | 2 × 10$^{-6}$ | 1400 | 2 hr |
| TiN/ScN Superlattice[34] | ~10$^{-4}$ | 1000 | 12 hr |
| TiN nanostructure[54] | x | 800 | 8 hr |
| HfN nanostructure[58] | ~ 10$^{-5}$ | 900 | 40 min |
| This work | 500 | 800 | 10 hr |

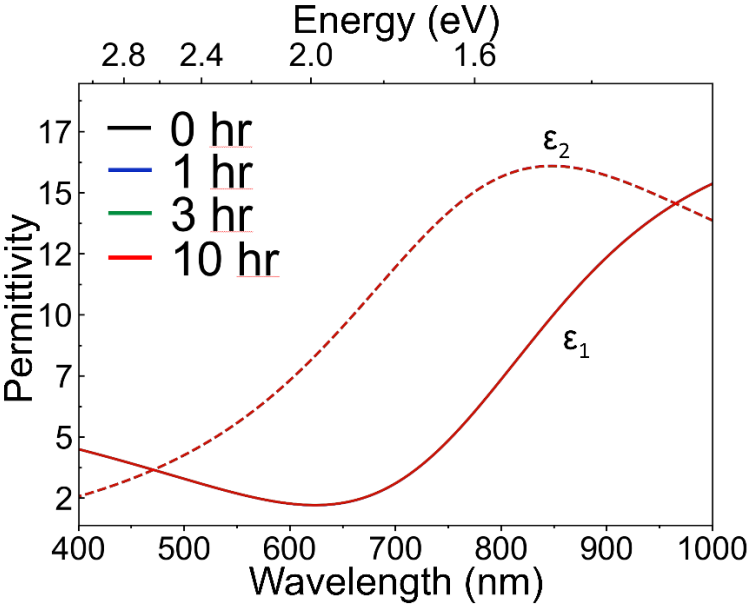

**Figure S11. Out-of-plane permittivity of the TiN/hBN superlattice during annealing.** The real (solid) and imaginary (dashed) parts of the effective permittivity of the TiN/hBN superlattice upon annealing at 800 °C for varying amounts of time. The permittivity was measured using the Parks Systems ellipsometer.

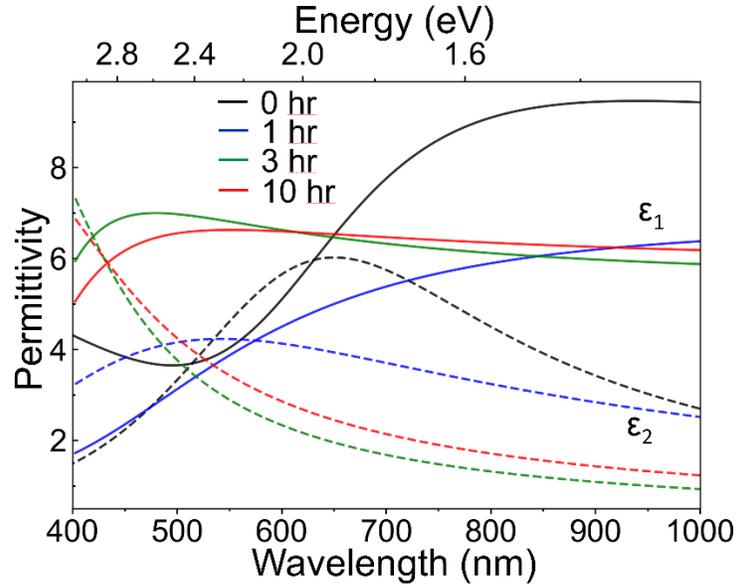

**Figure S12. Out-of-plane permittivity of the TiN/Al$_2$O$_3$ superlattice during annealing.** The real (solid) and imaginary (dashed) parts of the effective permittivity of the TiN/Al$_2$O$_3$ superlattice upon annealing at 800 °C for varying amounts of time. The permittivity was measured using the Parks Systems ellipsometer.

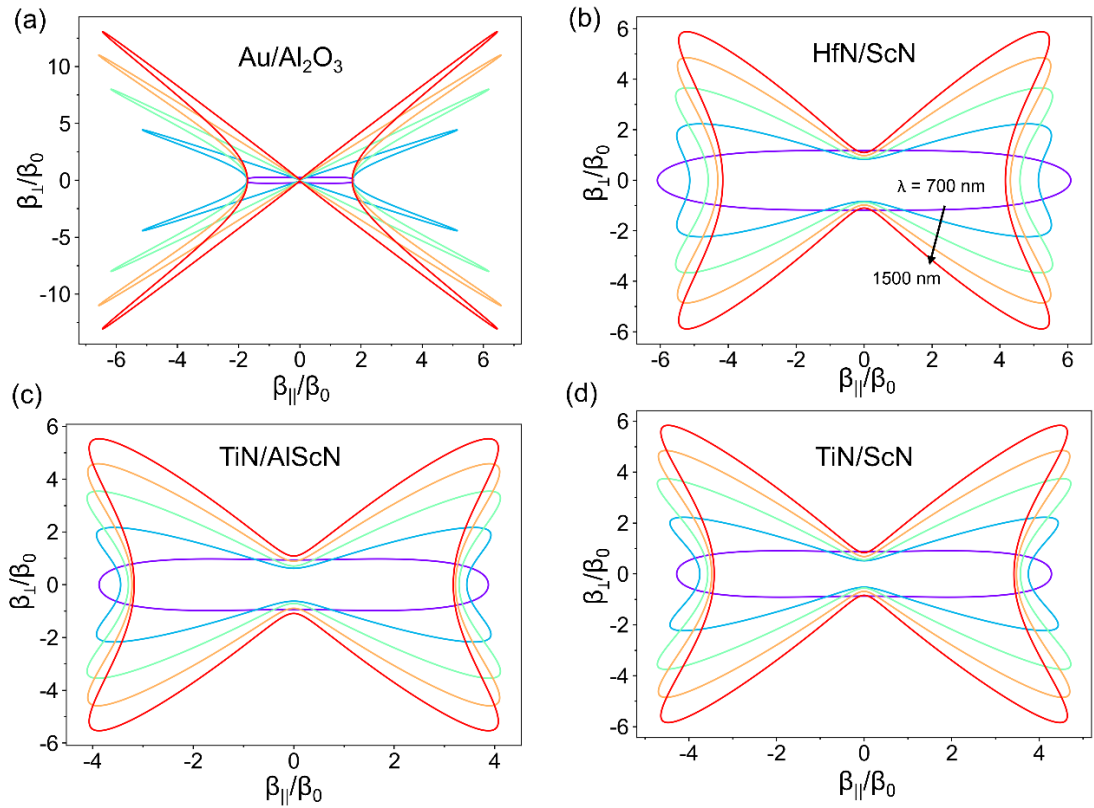

**Figure S13. Dispersion relation of hyperbolic superlattices in literature.** The dispersion relations of **(a)** Au/Al$_2$O$_3$, **(b)** HfN/ScN, **(c)** TiN/AlScN, and **(d)** TiN/ScN at wavelengths from 700 nm to 1500 nm at 200 nm

increments. The geometries of the superlattices have been set such that they all become hyperbolic at 700 nm, and the dispersion relation was calculated by including the effects of loss in the media[21].

| Substrate | c-plane Sapphire | hBN |
|---|---|---|
| Epsilon near zero (ENZ) point (eV) | 2.45 | 2.15 |
| Target | TiN (99.5%) | TiN (99.5%) |
| Base pressure (Torr) | $4\times 10^{-8}$ | $4\times 10^{-8}$ |
| Growth pressure (mTorr) | 3 | 3 |
| Ar (sccm)/$N_2$ (sccm) flow rate | 12:0 | 12:0 |
| Applied Power (W) | 110 | 110 |
| Growth temperature (°C) | 800 | 800 |

**Table S2. Material properties and growth parameters of TiN films by RF sputtering.** The ENZ property of the TiN film can be varied with different substrates. We sputtered quasi-epitaxial TiN film on Sapphire and hBN substrate with RF magnetron sputtering at a chamber pressure of approximately $10^{-8}$ torr and at a substrate temperature of 800 °C. The detail growth parameters show in following Table S1.